\documentclass{article}
\usepackage{waspaa21,amsmath,graphicx,svg,subcaption,bbm,caption}
\usepackage{amsfonts,amssymb,amsthm,amsmath,comment,color}
\usepackage{url,times}
\usepackage{microtype}
\usepackage{relsize}
\usepackage{subcaption}

\usepackage{cleveref}

\newcommand{\smallodot}{\odot}

\title{Weakly Supervised Source-Specific Sound Level Estimation in Noisy Soundscapes}
\name{
      Aurora Cramer, $^{1}$
      Mark Cartwright,$^{2}$
      Fatemeh Pishdadian,$^{3}$
      Juan Pablo Bello$^{1}$\thanks{Please direct correspondence to \protect\url{jtc440@nyu.edu}. \newline This work is partially supported by National Science Foundation award 1633259 (BIRDVOX) and award 1544753 (SONYC).}
      }
\address{$^1$ New York University, Music and Audio Research Laboratory, New York, NY, USA \\
         $^2$ New Jersey Institute of Technology, Newark, NJ, USA\\
         $^3$ Northwestern University, Evanston, IL, USA\\
}

\begin{document}
\ninept
\maketitle
\begin{abstract}
While the estimation of what sound sources are, when they occur, and from where they originate has been well-studied, the estimation of how loud these sound sources are has been often overlooked. Current solutions to this task, which we refer to as source-specific sound level estimation (SSSLE), suffer from challenges due to the impracticality of acquiring realistic data and a lack of robustness to realistic recording conditions. Recently proposed weakly supervised source separation offer a means of leveraging clip-level source annotations to train source separation models, which we augment with modified loss functions to bridge the gap between source separation and SSSLE and to address the presence of background. We show that our approach improves SSSLE performance compared to baseline source separation models and provide an ablation analysis to explore our method's design choices, showing that SSSLE in practical recording and annotation scenarios is possible.
\end{abstract}
\section{Introduction}
\label{sec:intro}
The field of machine listening is concerned with addressing the perception and characterization of acoustic scenes and their constituent sound events. Sound event recognition (SER) is one of the fundamental problems of machine listening, which is concerned with \emph{what} sound sources are present in a soundscape and when they occur \cite{virtanen2018computational}. Sound source localization is another fundamental problem concerned with from \emph{where} sound sources originate \cite{adavanne2018sound, madhu2008acoustic}. Both of these problems are well studied. Sound level estimation for specific sources, referred to henceforth as \emph{source-specific sound level estimation (SSSLE)}, is concerned with \emph{how loud} sound sources are, but is not well-studied in the case of sources in polyphonic soundscapes.

Despite being under-studied, SSSLE has many important yet often unrealized applications. In urban noise pollution monitoring, SSSLE could estimate the loudness of specific sound sources (e.g., traffic or construction equipment) to aid in noise mapping and enforcement \cite{DI2018115, GLOAGUEN2019229, gloaguen:hal-02285320, gloaguen:hal-01375796}. In intelligent audio production, SSSLE is used to determine the relative gain of instruments in audio mixes and inform automatic mixing systems that mimic audio engineers \cite{Scott11automaticmultitrack, ward2017estimating}. SSSLE could also be used to aid in distance estimation for diverse source localization applications such as wildlife monitoring \cite{lostanlen2019robust} and sound awareness technology for the hearing impaired \cite{jain2015head}.

%threat detection in autonomous vehicles,, and smart home technology
%Existing SSSLE approaches generally suffer from a few fundamental problems. Measuring sound levels for specific sources is often difficult or impossible in realistic (e.g. noisy and polyphonic) settings such as city streets. This makes it difficult to collect reliable ground truth for developing and evaluating SSSLE systems. Source separation based approaches can be applied, even when isolated sources are not available, However, these methods typically do not directly address SSSLE in environments with background noise. Thus, we posit that \emph{bridging the gap between source separation and SSSLE} and \emph{accounting for background} are important steps to move the field of SSSLE forward.

Existing SSSLE approaches suffer from the challenge of collecting reliable ground truth for developing and evaluating methods. Measuring source-specific sound levels for ground truth data is nontrivial in realistic (e.g., noisy and polyphonic) settings such as city streets. While source separation techniques can help to isolate sources in mixtures, these techniques do not perform well in background noise and out-of-vocabulary sources nor formulated to directly address SSSLE.

%In this work, we propose a framework for SSSLE that can be trained using either isolated sources, frame-level source presence annotations, or clip-level source present annotations. In particular, this framework addresses SSSLE by both \emph{bridging the gap between source separation and SSSLE} and \emph{accounting for background}. We evaluate our proposed models on a synthetic dataset of urban sound sources with realistic noise, and find that they outperform baseline source separation methods. Finally, ablation analyses are provided to investigate the effect of design choices in our framework.
In this work, we propose a framework for SSSLE that not only can be trained with just clip-level source presence annotations but also \emph{bridges the gap between source separation and SSSLE} and \emph{accounts for background}, both of which we posit are important steps to advance SSSLE. We evaluate our proposed models on a synthetic dataset of urban sound sources with realistic noise and out-of-vocabulary sources. We find that they outperform baseline source separation methods, and an ablation analysis investigates the effect of design choices in our framework. While our approach has been developed and evaluated using a synthetic dataset, one of its strengths is that it can be applied to real, noisy recordings with just clip-level annotations.
\section{Related Work}
\label{sec:related-work}
\subsection{Sound level estimation}
%Sound level estimation of mixtures and single sources is well-established in the space of acoustics \cite{kinsler1999fundamentals}. 
Typically, sound levels are measured for an entire audio signal, regardless of the sources, and can be measured at short-term scales or in an integrated fashion considering the entire signal. %Amplitude level measurements like \emph{RMS (root-mean square)} are often the most straightforward way of characterizing signal energy. %However, since digital audio signals are scaled depending on the recording device, 
For digital audio signals, full-scale sound level measures such as \emph{dBFS (decibels relative to full scale)} \cite{kinsler1999fundamentals} are typically used and can be calibrated to estimate sound pressure levels. 
In contrast, loudness measures like \emph{LUFS (loudness units relative to full scale)} apply perceptual weighting as well as silence gating mechanisms to provide sound level estimates more closely related to perceived loudness \cite{grimm2010toward}.

% In the area of speech enhancement, sound level estimation for background noise has been well studied. These established methods, which assume a psuedo-stationary model of background noise, estimate background power statistics to design time-varying filters to remove background from a speech recording \cite{ephraim1983speech, ephraim1984speech, ephraim1985speech}.
%Talk about old acoustics techniques, and asking people to turn off sources to measure source sound levels

Prior research in SSSLE has primarily focused on estimating traffic sound levels in noise monitoring and the estimation of isolated instrument track levels in automatic mixing systems. Traditional approaches to SSSLE require access to isolated sources \cite{Scott11automaticmultitrack, don1985road, perez2009automatic}. However, more recent approaches have used fully-supervised source separation techniques to perform well in the presence of other known, modeled sources \cite{GLOAGUEN2019229, gloaguen:hal-02285320, gloaguen:hal-01375796, ward2017estimating}.

%Within the scope of existing SSSLE literature, traffic sound level estimation has seen a fair amount of attention. Traditional methods take a lot of care in the data acquisition process to collect mostly isolated examples of traffic, but have no method for estimating traffic noise in the presence of other sounds \cite{don1985road}. More recent methods have used source separation algorithms such as non-negative matrix factorization (NMF) to estimate traffic sound-levels \cite{GLOAGUEN2019229, gloaguen:hal-02285320, gloaguen:hal-01375796}. In automatic mixing, while sound level estimation is typically applied to systems that have access to isolated tracks \cite{perez2009automatic, Scott11automaticmultitrack}, recent methods employ various out-of-the-box source separation methods for sound level estimation \cite{ward2017estimating}.

Using source separation \cite{vincent2018audiosourceseparation} to isolate a specific source and directly measure its sound levels is a straightforward approach to SSSLE. If we have a perfect source separation system (assuming no scale normalization), then we can perfectly estimate the sound level of each source. Time-frequency masking based deep learning methods are among the most popular source separation approaches, particularly because of the flexibility afforded by modern deep learning frameworks. However, while these methods perform fairly well for source separation, they suffer the drawback of requiring time-frequency-level ground-truth obtained from isolated recordings. Since obtaining isolated recordings in realistic soundscapes is generally impractical or impossible, training these models on realistic soundscapes can prove difficult \cite{salamon2017scaper}. Fortunately, recent methods have addressed this impracticality by training models using weaker supervision, via clip-level or frame-level annotations \cite{pishdadian2020finding}. We therefore restrict the scope of this paper to these methods.

\subsection{Discriminator-based weakly supervised source separation}
\label{sec:weakly-supervised}
Discriminator-based weakly supervised methods for source separation train a model to separate sources in a mixture using a classifier to critique the separated sources \cite{pishdadian2020finding, kong2018joint, kong2019sound} using only clip-level or frame-level annotations as well as \emph{energy consistency} between the mixture and the reconstructed sources \cite{pishdadian2020finding}. More recent methods leverage unsupervised methods involving learning to mix and separate mixtures \cite{wisdom2020mixit, kong2020source}, though we restrict the scope of this work to weakly supervised methods. We build upon on the framework presented by Pishdadian et al. \cite{pishdadian2020finding}, focusing on the clip-level scenario since it reflects practical SSSLE settings, and describe how \emph{energy consistency loss} and \emph{classification loss} enable weak supervision.

Formally, let $\mathbf{X} \in \mathbb{R}^{F \times T}_{\ge 0}$ be the time-frequency magnitude representation for a sound mixture and let $\mathbf{y} = [y_1, \ldots, y_C]^\intercal \in \{0, 1\}^C$ be the clip-level class presence vector. In this framework, a source separation network parameterized by $\theta$ takes $\mathbf{X}$ as input and produces a time-frequency mask $\mathbf{\hat{M}}_i = \mathbf{f}_{i, \theta}(\mathbf{X}) \in [0, 1]^{F \times T}$ for class $i$, resulting in the source estimate $\mathbf{\hat{S}}_i = \mathbf{\hat{M}}_i \smallodot \mathbf{X}$.

An \emph{energy consistency} loss $\mathcal{L}_{\text{mix}}$ is applied to ensure that estimated \emph{active sources} (classes with $y_i = 1$) contain the same energy as the mixture, and that estimated \emph{inactive sources} (classes with $y_i = 0$) contain no energy. Let $\mathcal{A}_{+}$ be the ground-truth set of active sources in the clip. We define the energy consistency residual for active sources, $\mathbf{R}_{\text{active}}  =  \mathbf{X} - \sum_{i \in \mathcal{A}_{+}} \mathbf{\hat{S}}_i$, and the energy consistency residual for inactive sources, $\mathbf{R}_{\text{inactive}} =  \sum_{j \not\in \mathcal{A}_{+}} \mathbf{\hat{S}}_j$. Then we have:
\begin{align}
    \mathcal{L}_{\text{mix}} & = \mathcal{L}_{\text{mix-active}} + \mathcal{L}_{\text{mix-inactive}} \\
     & = \frac{1}{TF} \left\|\mathbf{M}_E \smallodot \mathbf{R}_{\text{active}}\right\|_1 +
    \frac{1}{TF} \left\|\mathbf{M}_E \smallodot \mathbf{R}_{\text{inactive}} \right\|_1
%    \mathcal{L}_{\text{mix-active}} & = \frac{1}{TF} \left\|\mathbf{M}_E \smallodot \mathbf{E}_{\text{active}}\right\|_1 \\
%    \mathcal{L}_{\text{mix-inactive}} & = \frac{1}{TF} \left\|\mathbf{M}_E \smallodot \mathbf{E}_{\text{inactive}} \right\|_1
\end{align}
 where $\mathbf{M}_E \in \{0, 1\}^{F \times T}$ is a masking matrix that zeros out frames containing less than 1\% of the maximum frame energy in the clip. 
 By applying this mask, we only train on audio containing salient activity; whenever we normalize by $T$, we actually normalize by the number of salient frames, but omit this distinction for brevity. Additionally, we use $\| \cdot \|_1$ to indicate the element-wise $\ell_1$ norm; that is, $\|\mathbf{X}\|_1 = \sum_{i, j} |X_{ij}|$.
 
A classifier network then independently takes in each estimated $\mathbf{\hat{S}}_i$ as input, which should detect only the presence of class $i$ when indeed present. We also expect that when given the mixture $\mathbf{X}$ as input, the classifier's estimates for each class should match the ground-truth. Specifically, we apply the \emph{classification loss}:
\begin{align}
    \mathcal{L}_{\text{cls}} & = \mathcal{L}_{\text{cls-mix}} + \sum_{i} \mathcal{L}_{\text{cls-mix}, i}
\end{align}
where $\mathcal{L}_{\text{cls-mix}}$ indicates binary cross-entropy between clip-level mixture targets $y_i$ and mixture classifier predictions $\hat{y}_i^{(\text{ mix})}$ and $\mathcal{L}_{\text{cls-mix}, i}$ indicates binary cross-entropy between a modified target $\mathbf{y} \smallodot \mathbf{e}_i$ (where $\mathbf{e}_i$ is a canonical basis vector) and the clip-level classifier prediction for the separated source for class $i$. Note that when training the separator, $\mathcal{L}_{\text{cls-mix}}$ only has an effect if also optimizing the classifier. However, there is evidence that independently training the classifier and fixing the weights prior to training the separator results in improved separation \cite{pishdadian2020finding}; therefore, we follow this procedure.
The final loss for this framework is then:
\begin{equation}
\mathcal{L}_{\text{weak}} = \alpha \mathcal{L}_{\text{mix}} + \mathcal{L}_{\text{cls}}
\end{equation}
where $\alpha$ is a hyperparameter determining the relative importance of each term.
While this framework affords us the ability to train source separation models with more easily attainable annotations, which can make SSSLE more practical, it does not directly address SSSLE nor does it account for the presence of background noise or out-of-vocabulary events. Therefore, we propose an augmented framework to address these fundamental concerns.
\section{Source-Specific Sound Level Estimation}
\label{sec:methods}
\subsection{Bridging source separation and SSSLE}
The energy consistency terms in the loss for weakly supervised source separation are in the form $\frac{1}{TF} \|\mathbf{R}\|_1$. However, this term enforces consistency at the time-frequency scale, whereas we are ultimately interested in estimating the sound level. Note that the sound level can be estimated from either the energy spectrum or from the total energy of the signal. However, the classifier requires a time-frequency input to properly critique the separator estimates. Therefore, we can augment the energy consistency terms to enforce energy consistency at the spectrum level or the global energy level using the time-frequency error terms. For convenience, we call such augmentations \emph{sound level augmentations}. We propose an augmentation to this loss of the form:
\begin{align}
    \frac{1}{T F_\text{fb}} \| h_\Phi(\mathbf{R}) \|_1 &= \frac{1}{T F_\text{fb}} \| \mathbf{B}_L \mathbf{A} \mathbf{R} \mathbf{B}_R \|_1 \label{eq-general-consistency}
\end{align}
where $\mathbf{A} \in \mathbb{R}_{\ge 0}^{F_{\text{fb}} \times F}$ is a filter bank matrix with $F_{\text{fb}}$ frequency bands, $\mathbf{B}_L \in \mathbb{R}_{\ge 0}^{F_{\text{out}} \times F_{\text{fb}}}$ is a frequency aggregation matrix producing $F_{\text{out}}$ frequency bands, $\mathbf{B}_R \in \mathbb{R}_{\ge 0}^{T \times T_{\text{out}}}$ is a temporal aggregation matrix producing $T_{\text{out}}$ frames, and $\Phi = (\mathbf{A}, \mathbf{B}_L, \mathbf{B}_R)$. This parameterization allows energy error to be aggregated to enforce energy consistency at different time-frequency resolutions. $\mathbf{A}$ is used to implement a filter bank transformation (in the frequency domain) that can be applied to emphasize reconstruction in perceptually relevant frequency bands. In particular, we consider a mel frequency filter bank $\mathbf{A}_\text{mel}$. If no filter bank is used, $\mathbf{A} = \mathbf{A}_\text{linear} = \mathbf{I}_{F}$. We consider the following key choices of $\mathbf{B}_L$ and $\mathbf{B}_R$:
\begin{itemize}
    \item \emph{time-frequency energy consistency}: $\mathbf{B}_L = \mathbf{I}_{F}, \mathbf{B}_R = \mathbf{I}_T$
    \item \emph{spectrum energy consistency}: $\mathbf{B}_L = \mathbf{I}_{F}, \mathbf{B}_R = \mathbf{1}_T$
    \item \emph{global energy consistency}: $\mathbf{B}_L = \mathbf{1}_{F}^\intercal, \mathbf{B}_R = \mathbf{1}_T$
\end{itemize}
%As shorthand, we denote $\Phi_{\text{time-freq}, \mathcal{F}}, \Phi_{\text{spectrum}, \mathcal{F}}, \Phi_{\text{global}}$ as the parameterization for time-frequency, spectrum, and global energy consistency, respectively, where $\mathcal{F} \in \{\text{linear}, \text{mel}\}$ corresponds to the choice of the filter bank matrix $\mathbf{A}_\text{linear}$ or $\mathbf{A}_\text{mel}$.
While using only the global energy consistency configuration or even spectrum energy consistency configuration would most closely match the end goal of sound level estimation, we would lose the time-frequency structure afforded by the time-frequency energy consistency configuration. The classifier still enforces time-frequency structure by classifying the estimated spectrograms; however, without the time-frequency consistency loss there is less incentive for the estimated sources to resemble the original mixture.
Therefore, we propose enforcing mean energy consistency across multiple time-frequency resolutions. Specifically, for a set of parameterizations $\mathcal{P} = \{\Phi_1, \ldots, \Phi_P\}$, we can take the mean energy consistency:
\begin{equation}
\frac{1}{|\mathcal{P}|} \sum_{\Phi \in \mathcal{P}} \frac{1}{T F_\text{fb}} \| h_\Phi(\mathbf{R}) \|_1
\end{equation}
With this method, we can easily enforce energy consistency at various resolutions to retain time-frequency structure while directly addressing sound level estimation. We therefore choose $\mathcal{P}_\text{all, mel}= \{\Phi_{\text{time-freq, mel}}, \Phi_{\text{spectrum, mel}}, \Phi_{\text{global}} \}$ to incorporate multi-resolution structure and choose $\mathbf{A}_\text{mel}$ focus on perceptually relevant frequency bands with mel frequency filter banks.
However, for global energy consistency we always use $\mathbf{A}_\text{linear}$ to avoid redundancy.

\subsection{Accounting for background}
The energy consistency term must be modified in the presence of background noise and out-of-vocabulary events, since the sum of the sources of interest will not in general result in the mixture. For convenience, we call such augmentations \emph{background augmentations}. A straightforward solution is to allow for the sum of source energy to underestimate the mixture energy. To achieve this, we can again modify Eq. \ref{eq-general-consistency} using an asymmetric margin parameterized by $\varepsilon \ge 0$:
\begin{align}
    \|\mathbf{R}\|^{(\text{asym}, T,F, \varepsilon)}_1 & = \left[\left\|\left[\mathbf{R}\right]_+\right\|_1 - TF \varepsilon \right]_+ + \left\| \left[- \mathbf{R} \right]_+\right\|_1
\end{align}
where $[\ \cdot \ ]_+$ indicates element-wise half-wave rectification. Note that $\varepsilon$ specifies the allowable mean energy per time-frequency bin and that when $\varepsilon = 0$, $\|\mathbf{R}\|^{(\text{asym}, \varepsilon)}_1 = \|\mathbf{R}\|_1$. A reasonable choice for $\varepsilon$ is the mean energy margin estimated from the training set. 
We then have a residual signal containing only background and out-of-vocabulary sources, and therefore the classifier should indicate that no in-vocabulary sources are present. To the residual spectrogram estimate $\mathbf{\hat{S}}_{\text{bkgr}} = \left[1 - \sum_{i} \hat{\mathbf{M}}_i\right]_+ \smallodot \mathbf{X}$, we apply the loss: \begin{equation}
    \mathcal{L}_{\text{cls-bkgr}} = \sum_{i} H\left(0, \hat{y}_i^{(\text{bkgr})}\right)
\end{equation} where $\hat{y}_{i}^{(\text{bkgr})}$ is the clip-level classifier prediction for class $i$ for the residual spectrogram  $\mathbf{\hat{S}}_{\text{bkgr}}$ and $H$ is the binary cross-entropy function.
%where $\hat{y}_{i, t}^{(\text{clip, bkgr})}$ is the classifier prediction for class $i$ at time $t$ for the residual spectrogram  $\mathbf{\hat{S}}_{\text{bkgr}}$.
By allowing for a residual signal and by enforcing that it does not contain any in-vocabulary sound sources, we can handle additive background noise and out-of-vocabulary sources in a principled way.

\subsection{Putting it all together}
Combining the aforementioned augmentations with the approach detailed in Section \ref{sec:weakly-supervised}, we have the components to build our framework for SSSLE. Our SSSLE model optimizes the loss:
\begin{align}
    \mathcal{L}^{\mathcal{P}}_{\text{weak, sssle}} & = \frac{\alpha}{|\mathcal{P}|} \sum_{\Phi \in \mathcal{P}} \mathcal{L}^{\Phi}_{\text{mix, sssle}} + \mathcal{L}_{\text{cls, sssle}}
\end{align}
where we have
\begin{align}
    \mathcal{L}^{\Phi}_{\text{mix, sssle}} & = \mathcal{L}^{\Phi}_{\text{mix-active, sssle}} + \mathcal{L}^{\Phi}_{\text{mix-inactive, sssle}} \\
    \mathcal{L}^{\Phi}_{\text{mix-active, sssle}} & =    \frac{1}{T F_\text{fb}} \|h_\Phi(\mathbf{M}_E \smallodot \mathbf{R}_{\text{active}})\|^{(\text{asym}, T,F_\text{fb}, \varepsilon)}_1 \\
    \mathcal{L}^{\Phi}_{\text{mix-inactive, sssle}} & = \frac{1}{T F_\text{fb}} \left\|h_\Phi\left(\mathbf{M}_E \smallodot \mathbf{R}_{\text{inactive}}\right)\right\|_1 \\
    \mathcal{L}_{\text{cls, sssle}} & = \mathcal{L}_{\text{cls-mix}} + \sum_i \mathcal{L}_{\text{cls-mix}, i} + \beta \mathcal{L}_{\text{cls-bkgr}}
\end{align}
where $\beta \in \{0, 1\}$. When $\beta = 1$, $\mathcal{L}_{\text{cls-bkgr}}$ is included in the loss, which ensures that the residual signal (containing the background) does not contain the presence of any of the in-vocabulary classes.
%Since the fully supervised source separation model directly estimates each isolated source independently, the augmentations for handling background are not necessary. However, we can still apply the energy consistency augmentations. Therefore, when training SSSLE models with TF-bin-level annotations, the model optimizes the loss $\mathcal{L}_\text{strong, sssle}^{\Phi}$:
%\begin{equation}
%    \mathcal{L}_\text{strong, sssle}^{\Phi} = \frac{1}{|\mathcal{P}|} \sum_{\Phi \in \mathcal{P}} \frac{1}{T_\text{out} F_\text{out}} \sum_{i} \left\|h_\Phi\left(\mathbf{M}_E \smallodot \mathbf{W}_i \smallodot \mathbf{R}_i \right)\right\|_1
%\end{equation}
With this framework, we can more directly address SSSLE within the weakly supervised source separation framework.

\section{Experimental Methods}

The procedure we use to evaluate our proposed methods is as follows. First, we generate a dataset of soundscapes containing sound events and varying levels of background noise, with accompanying clip-level source annotations. We then train models with our proposed loss augmentations, compare them to baseline source separation models, and perform an ablation analysis on our method's design choices. To evaluate source separation performance, we use scale-invariant signal-to-noise ratio (SI-SDR) \cite{le2019sdr} improvement with respect to the baseline of using the mixture as the estimate. The reconstructed audio signals are obtained by taking the ISTFT of the masked spectrogram with the mixture phase. To evaluate sound level estimation, we use the absolute error with respect to dBFS. In the proceeding sections, we describe the components of our experiments.

\subsection{Dataset creation}
\label{sec:experiments-datasets}

To train and evaluate these models, we use the synthetic dataset of urban soundscapes\footnote{Thanks to Gordon Wichern at Mitsubishi Electric Research Laboratory for their correspondence and use of their data.} used by Pishdadian et al.~\cite{pishdadian2020finding}. This dataset contains 4 second synthetic mixtures sampled at 16 kHz containing foreground events from UrbanSound8K \cite{salamon2014dataset}, specifically from the 5 classes \emph{car horn}, \emph{dog bark}, \emph{gun shot}, \emph{jackhammer}, and \emph{siren}. 
%50,000, 10,000, and 10,000 examples are generated for the training, validation, and testing sets, respectfully. 
Folds 1--6, 7--8, and 9--10 of UrbanSound8K are used to generate the training, validation, and testing subsets, respectively, with 50k, 10k, and 10k examples each. Each soundscape contains a number of events sampled from a zero-truncated Poisson distribution ($\lambda=5$), with a uniformly random start-time and a uniformly chosen class.

To evaluate the methods in the presence of background noise and potentially out-of-vocabulary events, we add background noise to the base dataset at varying sound levels. For the background audio, we use audio clips recorded by the Sounds of New York City acoustic sensor network \cite{bello2019sonyc} that an urban sound tagging classifier identified as not containing urban sound classes. The classifier was trained on SONYC-UST-V1 \cite{cartwright2019sonycust} and uses OpenL3 embeddings \cite{cramer2019look} as input. We have released these background clips in the SONYC-Backgrounds\footnote{SONYC-Backgrounds: \protect\url{https://doi.org/10.5281/zenodo.5129078}} dataset.
%The lack of presence is determined by choosing presence probabilities below a chosen threshold (corresponding to a true-negative-rate of 70\% on the SONYC-UST-V1 test set).
We create a train/valid/test split by choosing disjoint sensors resulting in a roughly 60/20/20 split. For each example in the base dataset, we uniformly sample a 4-second segment of a background clip, which is subsequently LUFS-normalized mixed with the original example. We create datasets using background sound levels $\in \{-50, -20, 0\}$ dB LUFS, which we refer to as \emph{weak} background, \emph{moderate} background, and \emph{strong} background. We have made these datasets available for the sake of reproducability.\footnote{Soundscape data: \protect\url{https://doi.org/10.5281/zenodo.5123372}} 
\subsection{Baseline comparisons}
\label{sec:experiments-baseline}
For our baseline, we use the models without the augmentations outlined in Section \ref{sec:weakly-supervised}. We also compare with models individual and combined effects of sound level augmentations and background augmentations. %Note that we focus our results on models using clip-level since that is this is the most realistic scenario, and as we found that the patterns found within models trained with weak clip-level annotations are also reflected in models trained with other annotations. 
Our proposed models are trained to minimize $\mathcal{L}_\text{weak, sssle}^{\mathcal{P}}$, with $\mathcal{P} = \mathcal{P}_\text{all, mel}$ and $\beta = 1$. We choose $\alpha = 100$ to remain consistent with previous work \cite{pishdadian2020finding}. Classifiers are trained separately to minimize $\mathcal{L}_\text{cls-mix}$, with classifier weights are frozen when training the separation model. $\varepsilon$ is estimated using the empirical mean of time-frequency margins between active sources and mixtures from training examples. All separation models are trained with all levels of background noise (none, weak, moderate, and strong).
%, and are compared with respect to source separation performance and SSSLE performance.

%Finally, we compare training with frame-level annotations and TF-bin-level annotations to observe the gains afforded from using annotations with greater time and/or frequency resolution.

\subsection{Ablation experiments}
\label{sec:experiments-ablation}
We perform two ablation studies to explore the design choices in the sound level and background augmentations.
For the sound level augmentations, we look at all combinations of time-frequency, spectrum, and global energy consistency loss as well as the use of mel frequency bands. Since background noise presence is not central to these augmentations, we restrict our attention to models trained on noise-less mixtures. For the background augmentations, we disable residual background classification and remove the energy margin to explore their effect on model performance in background noise. 

\subsection{Training details}
\label{sec:experiments-training}

For the front-end to the models, we use a log-magnitude STFT with a DFT-size of 512 with 25\% overlap with a square-root Hann window applied. When mel frequency filter banks are used, we use 40 bands. For the source separation models, instead of the BLSTM used by Pishdadian et al.  \cite{pishdadian2020finding}, we use a variant of the popular UNet model used by Kong et al. \cite{kong2020source}, removing the conditioning mechanism and increasing the number of outputs to the number of sound sources.
%We apply a natural log to the spectrogram as preprocessing to the separation models. 
For the classification models, we use the linear-frequency CRNN model used by Pishdadian et al. \cite{pishdadian2020finding}, training it on the base dataset with no added background.
All models are trained with the Adam optimizer, with initial learning rate $10^{-4}$ and batch size 8, for up to 50 epochs using early stopping on the validation set with a patience of 5 epochs. Our training code is available on our GitHub repository\footnote{Code: \protect\url{https://github.com/sonyc-project/weakly-supervised-sssle}}.

%\subsection{Evaluation}
%\label{sec:experiments-evaluation}
%To evaluate source separation performance, we use scale-invariant signal-to-noise ratio (SI-SDR) \cite{le2019sdr} improvement with respect to the baseline of using the mixture as the estimate. The reconstructed audio signals are obtained by taking the ISTFT of the masked magnitude spectrogram with the mixture phase. To evaluate sound level estimation, we consider the improvement of absolute error with respect to dBFS with respect to the same mixture baseline.

\section{Results and Discussion}
\subsection{Baseline comparison results}
\label{sec:results-results}
\begin{figure}[h]
     \centering
     \begin{subfigure}[b]{0.475\textwidth}
         \begin{subfigure}[b]{0.475\textwidth}
             \centering
             \includegraphics[width=\textwidth]{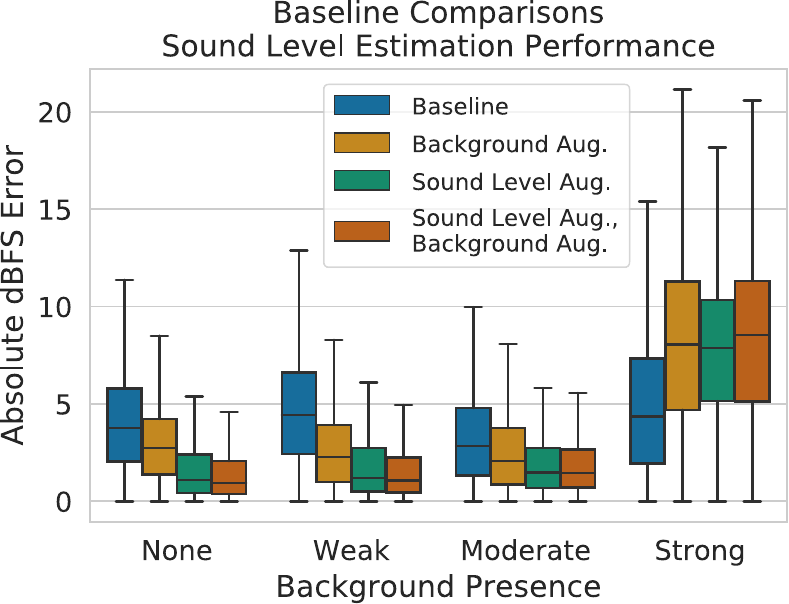}
         \end{subfigure}
         \begin{subfigure}[b]{0.475\textwidth}
             \centering
             \includegraphics[width=\textwidth]{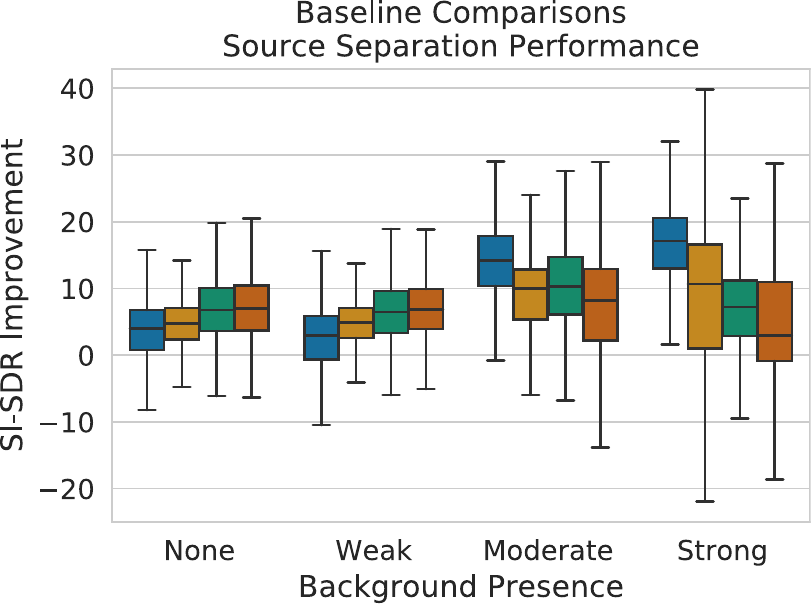}
         \end{subfigure}
         \caption{Baseline comparison results}
         \vspace{1.1em}
         \label{fig:baseline}
     \end{subfigure}
     \begin{subfigure}[b]{0.475\textwidth}
         \centering
         \begin{subfigure}[b]{0.475\textwidth}
             \centering
             \includegraphics[width=\textwidth]{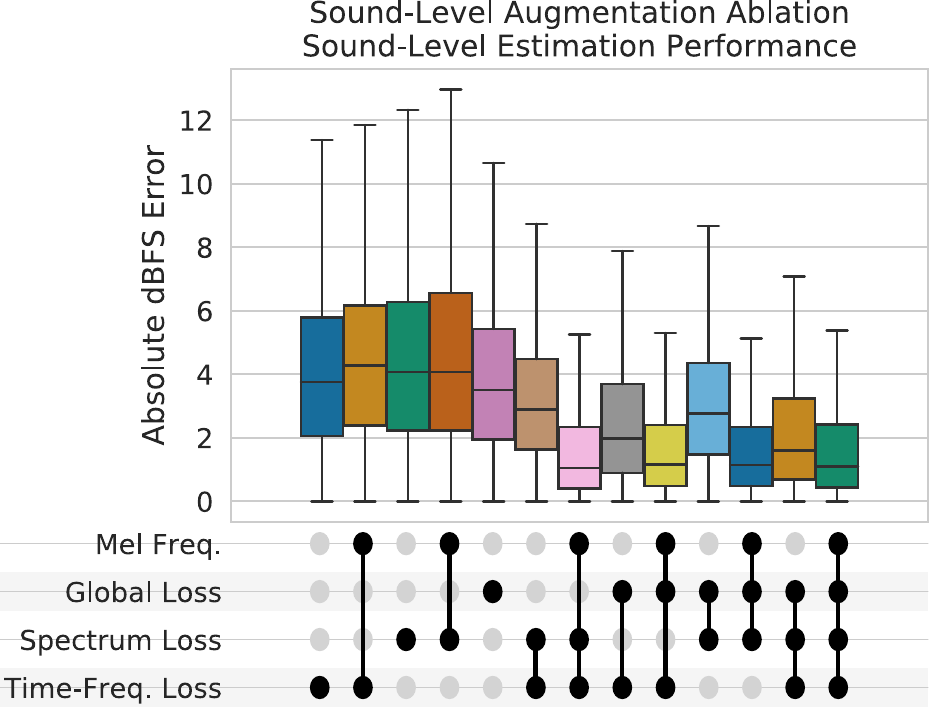}
             \label{fig:loudness-ablation}
         \end{subfigure}
         %\hfill
         \begin{subfigure}[b]{0.475\textwidth}
             \centering
             \includegraphics[width=\textwidth]{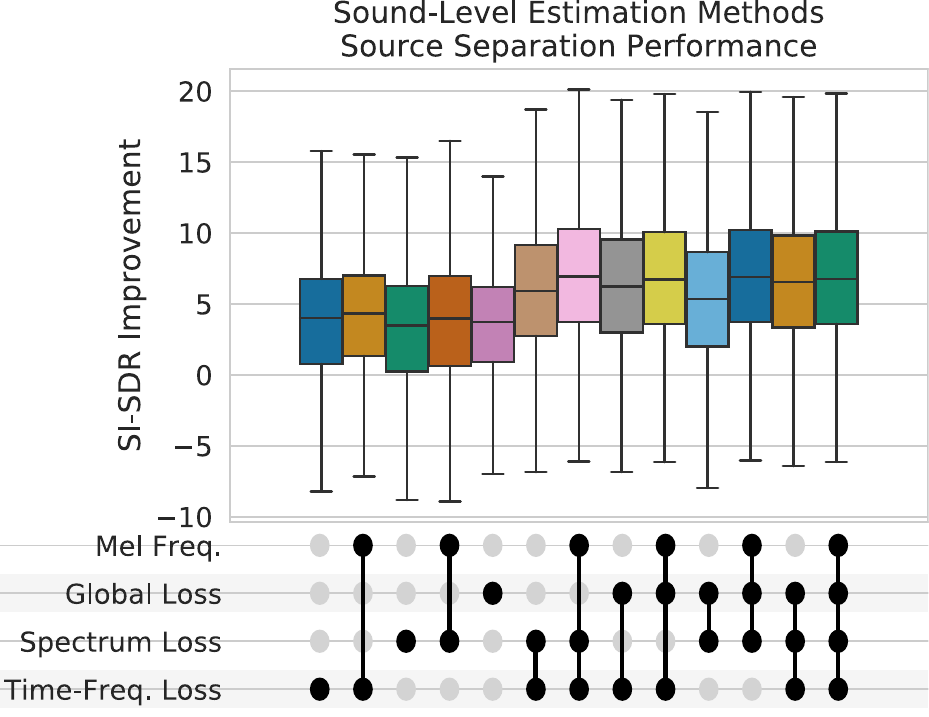}
         \end{subfigure}
         \caption{Sound level augmentation ablation study results}
         \vspace{1.1em}
         \label{fig:loudness-ablation}
    \end{subfigure}
    \begin{subfigure}[b]{0.475\textwidth}
         \centering
         \begin{subfigure}[b]{0.475\textwidth}
             \centering
             \includegraphics[width=\textwidth]{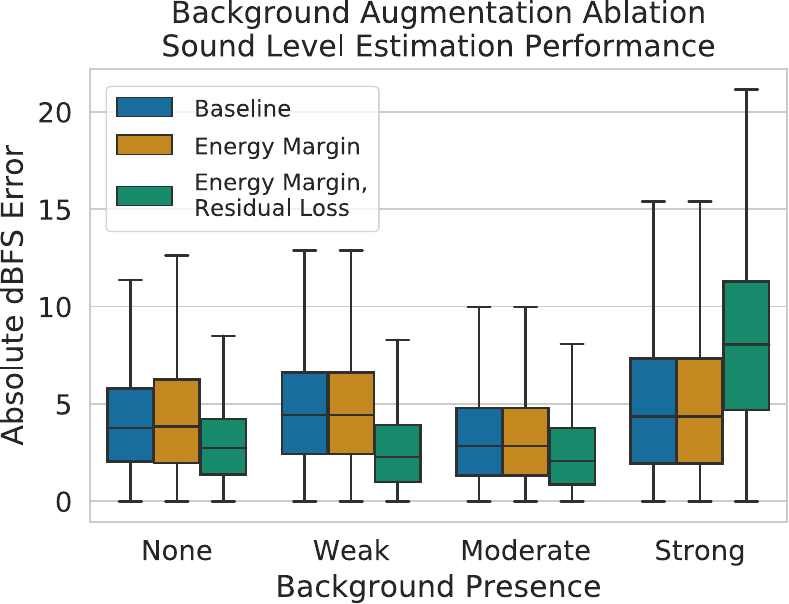}
         \end{subfigure}
         %\hfill
         \begin{subfigure}[b]{0.475\textwidth}
             \centering
             \includegraphics[width=\textwidth]{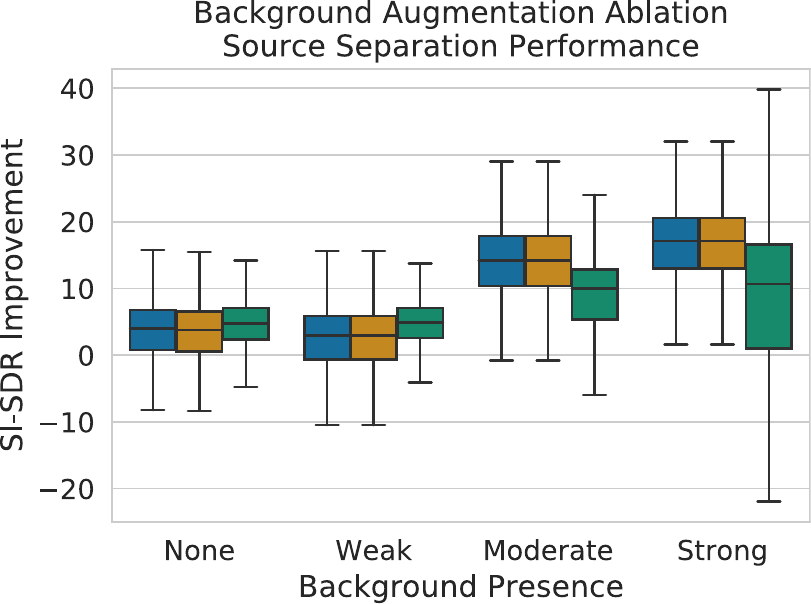}
         \end{subfigure}
         \caption{Background augmentation ablation study results}
         \label{fig:background-ablation}
    \end{subfigure}
     \caption{Boxplots of evaluation metrics across test examples (outliers omitted) for models in each set of experiments. (left): Sound level estimation performance w.r.t. absolute dBFS error. Lower is better. (right): Source separation performance w.r.t. SI-SDR improvement. Higher is better. }
    
\end{figure}
%\begin{figure}[h]
%     \centering
%     \begin{subfigure}[b]{0.23\textwidth}
%         \centering
%         \includegraphics[width=\textwidth]{baseline-dbfs_shared_leg.pdf}
%         \label{fig:baseline-dbfs}
%     \end{subfigure}
%     \begin{subfigure}[b]{0.23\textwidth}
%         \centering
%         \includegraphics[width=\textwidth]{baseline-sisdr_shared_leg.pdf}
%         \label{fig:baseline-sisdr}
%     \end{subfigure}
%     \caption{Boxplots of evaluation metrics across test examples (outliers omitted) for models in the baseline comparison experiment. (left): Sound level estimation performance w.r.t. absolute dBFS error. Lower is better. (right): Source separation performance w.r.t. SI-SDR improvement. Higher is better.}
%     \label{fig:baseline}
%\end{figure}
We see in Figure \ref{fig:baseline} that our proposed augmentations improve SSSLE performance and that using both sound-level and background augmentations provide the best performance.
We see significant performance improvement in background noise, though our approach still fails in high noise scenarios. This may be because with a sufficiently large margin, the mixture loss no longer provides enough structure to reconstruct sources in the mixture.
Source separation performance also improves in low noise conditions with our augmentations, though the baseline performs better with stronger noise, showing that our methods can improve source separation performance in some cases.
%As expected, using frame-level and TF-bin-level annotations further improve results.

\subsection{Ablation analysis}
\label{sec:results-ablation}

%\begin{figure}[h]
%     \centering
%     \begin{subfigure}[b]{0.23\textwidth}
%         \centering
%         \includegraphics[width=\textwidth]{sound-level-aug-dbfs_w_attrib.pdf}
%         \label{fig:loudness-ablation-dbfs}
%     \end{subfigure}
%     %\hfill
%     \begin{subfigure}[b]{0.23\textwidth}
%         \centering
%         \includegraphics[width=\textwidth]{sound-level-aug-sisdr_w_attrib.pdf}
%         \label{fig:loudness-ablation-sisdr}
%     \end{subfigure}
%     \caption{Boxplots of evaluation metrics across test examples (outliers omitted) for models in the ablation study for sound level augmentations. (left): Sound level estimation performance w.r.t. absolute dBFS error. Lower is better. (right):  Source separation performance w.r.t. SI-SDR improvement. Higher is better. }
%     \label{fig:loudness-ablation}
%\end{figure}

From the ablation analysis of the sound level augmentations shown in Figure \ref{fig:loudness-ablation}, we see that using multiple time-frequency resolutions improved sound level estimation, though global energy consistency helps less than spectrum energy consistency. This may indicate that while temporal aggregation helps, constraining frequency structure is still important. Additionally, using mel frequency scales in all cases is more effective. From the ablation analysis of the background augmentations shown in Figure \ref{fig:background-ablation}, introducing an energy margin and classifying the residual improve performance together.

\section{Discussion and future perspectives}
\label{sec:results-conclusion}
While the results are preliminary, we have shown that extending weakly supervised source separation methods to directly address sound level estimation and to handle background noise improves the use of such systems for SSSLE, showing great promise for the use of clip-level annotations for SSSLE in realistic recording scenarios.
While in this study we trained and evaluated using synthetic mixtures, real recordings of soundscapes with clip-level annotations could also be used. Evaluating SSSLE performance with a qualitative sound level metric like dBFS generally remains an open question, since obtaining ground truth sound level measurements is generally impossible in realistic scenarios. Subjective listening tests could be used to judge source-specific loudness, though this requires further development.
By showing that SSSLE is possible in realistic recording scenarios with only clip-level annotations, we hope to engender enthusiasm and research around moving forward SSSLE.

\bibliographystyle{IEEEtran}
\bibliography{main.bib}

\end{document}